\RequirePackage{fix-cm}
\documentclass[smallextended]{svjour3}       
\smartqed  
\usepackage{graphicx}
\usepackage{float}
\usepackage{subfig}
\usepackage{amsmath}
\usepackage{bm}
\usepackage{amsmath}
\usepackage{amssymb}
\usepackage{latexsym}
\usepackage{epsfig}
\usepackage{amsbsy}
\usepackage{array}
\usepackage{amssymb}
\usepackage{setspace}

\journalname{Plasmonics}
\begin{document}

\title{An ultrasensitive and multispectral refractive index sensor design based on quad-supercell metamaterials}



\author{Shuyuan Xiao \and Tao Wang($\boxtimes$) \and Yuebo Liu \and Xu Han \and Xicheng Yan}

\institute{Shuyuan Xiao \and Tao Wang($\boxtimes$) \and Xu Han \and Xicheng Yan
           \at Wuhan National Laboratory for Optoelectronics, Huazhong University of Science and Technology, Wuhan 430074, People's Republic of China \\
           \email{wangtao@hust.edu.cn}
           \and Yuebo Liu
           \at School of Information and Optoelectronic Science and Engineering, South China Normal University, Guangzhou 510006, People's Republic of China}

\date{Received: date / Accepted: date}

\maketitle

\begin{abstract}
Plasmonic metamaterials support the localized surface plasmon resonance (LSPR), which is sensitive to the change in the dielectric environment and highly desirable for ultrasensitive biochemical sensing. In this work, a novel design of supercell metamaterials of four mutually rotating split ring resonators (SRRs) is proposed, where simultaneous excitations of odd ($N=1$ and $N=3$) and even ($N=2$) resonance modes are realized due to additional asymmetry from the rotation and show insensitivity to two orthogonal polarizations. The full utilization of these three resonance dips show bright prospects for multispectral application. As a refractive index (RI) sensor, ultrahigh sensitivities $\sim1000$ nm/RIU for $LC$ mode ($N=1$) and $\sim500$ nm/RIU for plasmon mode ($N=2$) are obtained in the near infrared (NIR) spectrum.
\end{abstract}

\section{Introduction}\label{sec1}
Plasmonic metamaterials are artificially periodic structure of metallic nanoparticles much smaller than the corresponding operating wavelength\cite{soukoulis2007,shalaev2007,liu2009nm} and support the localized surface plasmon resonance (LSPR),
which comes from the collective oscillation of free electrons and induces the local electromagnetic field confinement\cite{hutter2004,tong2014}. A lot of novel electromagnetic phenomena with no analog in nature have been reported and employed to a great variety of potential applications such as electromagnetic cloaking\cite{schurig2006}, slow-light effect\cite{wu2011}, and molecular spectroscopy\cite{li2013}. It is also noted that the LSPR depends strongly on the dielectric environment and is sensitive to the change in refractive index (RI), which is highly desirable for ultrasensitive biochemical sensing\cite{liu2010infrared,jamali2014plasmonic,wang2015}. In 2009, a planar complementary metamaterial sensor based on plasmon coupling effects was experimentally demonstrated by Liu \emph{et al.}\cite{liu2009nl} and yielded a sensitivity of 588 nm/RIU. Since then, various biochemical sensors based on plasmonic metamaterials have been presented. For example, Cetin \emph{et al.}\cite{cetin2012} introduced a conducting metal layer underneath a Fano resonant asymmetric ring/disk plasmonic nanocavity system and increased the interaction volume of analytes and optical fields, and obtained a high refractive index sensitivity as large as 648 nm/RIU. Wu \emph{et al.}\cite{wu2014} designed a vertical nanostructure capable of lifting localized plasmon field off of the substrate, therefore effectively increased the sensing volume and predicted a sensitivity of about 800 nm/RIU. Unfortunately, these efforts share the common shortcoming of single operating wavelength, which greatly restricts their applications in practice. In many cases, a multispectral sensor is urgently needed since the ability to monitor multiple wavelengths allows one to correlate and study structural changes between different molecular regions and is critical to the accurate identification of biochemical molecules\cite{jiang2008,chen2012}.

Split ring resonator(SRR), first introduced by Hardy and Whitehead to realize the high $Q$ magnetic resonance in the frequency region 200-2000 MHz\cite{hardy1981}, is an analog of a $LC$-oscillator circuit where the crescent arc resembles the inductance ($L$) and the split serves as the capacitance ($C$). The noble metal based SRR supports the LSPR and is often adopted as the basic building block of plasmonic metamaterials\cite{zhang2013,wang2014jmo,wang2014jo,zhu2015}. As kind of a simple and elegant structure, the SRR has a group of tuneable parameters, the morphology, the size and periodicity and the LSPR is highly dependent on them. By adjusting these structure parameters, the LSPR has been respectively demonstrated in microwave\cite{smith2004,han2015}, infrared\cite{linden2004} and terahertz regions\cite{singh2009,singh2011,al-naib2015,wang2015design}, and those generated in the visible and near infrared (Vis-NIR) spectrum\cite{clark2007,xu2011,tobing2013,tobing2014,tobing2016} are of considerable interest in the field of ultrasensitive biochemical sensing since it is the exact portion of the spectrum most relevant to many biochemical molecules, such as proteins, glucose, and DNA. The fabrication of the corresponding 100-nm size or smaller SRRs, once considered as a key obstacle, has been realized by a robust electron beam lithography (EBL) with high contrast in recent years\cite{tobing2013,tobing2014,tobing2016}.

The feature that the inefficient excitation of resonance modes due to the strong polarization dependence once impedes the full utilization of the SRR, i.e., when the electric field is parallel to its split, only the two odd resonance modes ($N=1$ and $N=3$) are excited, while only the single even resonance mode ($N=2$) emerges when the electric field is perpendicular to the split\cite{clark2007,rockstuhl2006oe}. To address this concern, we propose a novel design in this work, supercell metamaterials consisting of four SRRs that are mutually rotated by 45$^{\circ}$ in each unit cell, where the odd and even resonance modes are simultaneously excited and show insensitivity to two orthogonal polarizations. The full utilization of these three resonance dips show bright prospects for multispectral application. As a RI sensor, all the three resonance dips show clearly observable shifts with the slight increase in the RI of the surrounding environment. Especially, ultrahigh RI sensitivities $\sim1000$ nm/RIU for $LC$ mode ($N=1$) and $\sim500$ nm/RIU for plasmon mode ($N=2$) in NIR are obtained simultaneously and reveal bright prospects in multispectral sensing.

\section{The quad-supercell metamaterial structure and numerical model}\label{sec2}
Fig. 1 schematically shows three quad-supercells of the SRRs with different mutual rotation angles $\varphi=0^{\circ}$, $45^{\circ}$ and $90^{\circ}$. Silver is adopted as the constituting metal of the SRRs due to its low optical loss and high plasma frequency. Each identical SRR has a size $s=100$ nm, a width $w=20$ nm, a thickness $h=30$ nm (thicker than the skin depth of silver $\sim22$ nm\cite{johnson1972}) and a split angle $\theta=30^{\circ}$, and the lattice constant is $a=180$ nm. The finite-difference time-domain method with the commercial software FDTD Solutions is employed here. The calculation step is fixed as 5 nm$\times$5 nm$\times$5 nm. The periodical boundary conditions with $p=500$ nm apply in both $x$ and $y$ directions, and perfectly matched layers in $z$ direction are utilized along the propagation of light. It is noted that only the $x$-polarization planewave is employed since the same results should be obtained with the $y$-polarization planewave due to obvious rotational symmetric relations among these three arrays. The substrate is quartz with a RI of 1.4 and frequency-dependent optical properties of silver are extracted from Palik's data\cite{palik1998}.
\begin{figure}[htbp]\label{fig:1}
\centering
\subfloat[]{
\includegraphics[scale=0.16]{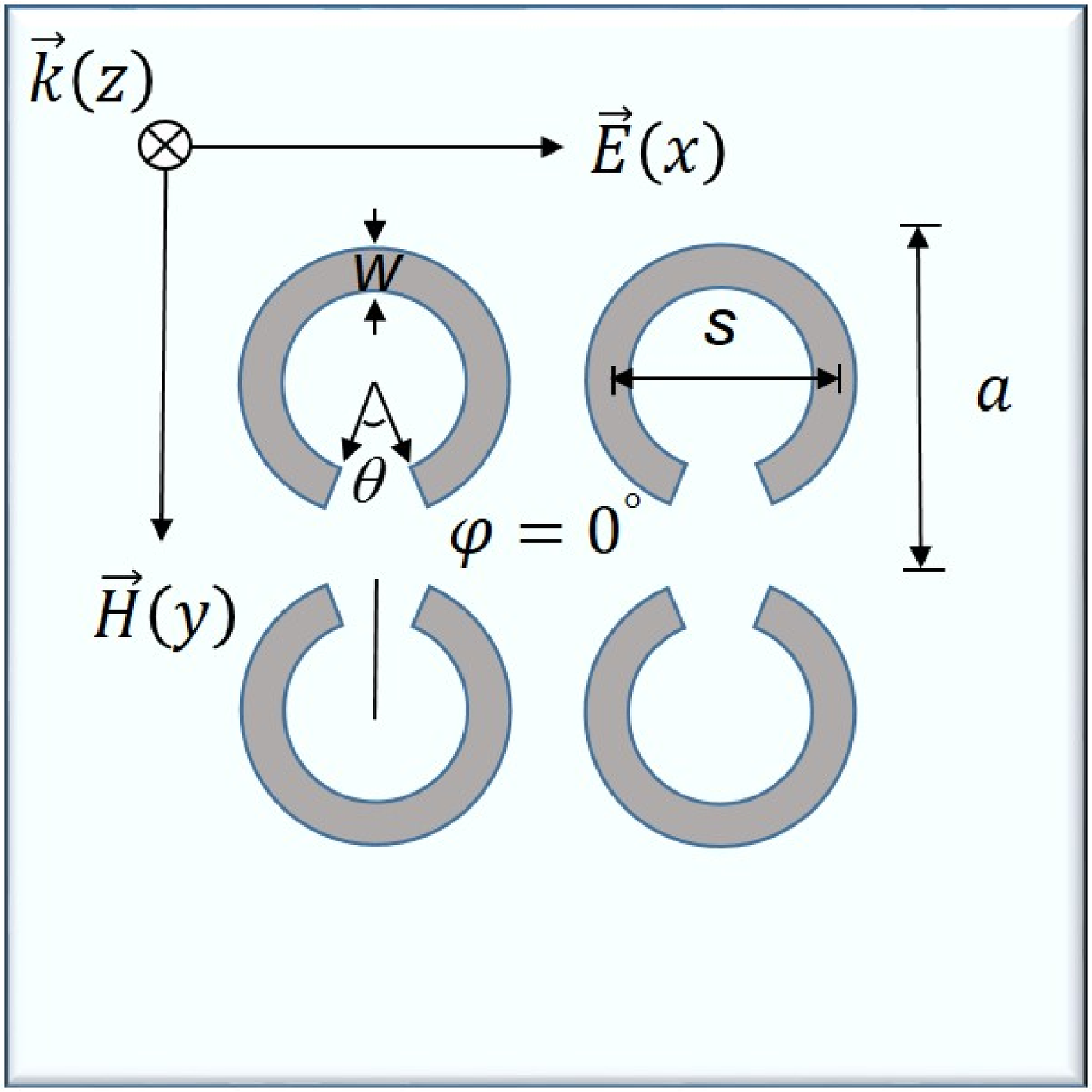}
}
\subfloat[]{
\includegraphics[scale=0.16]{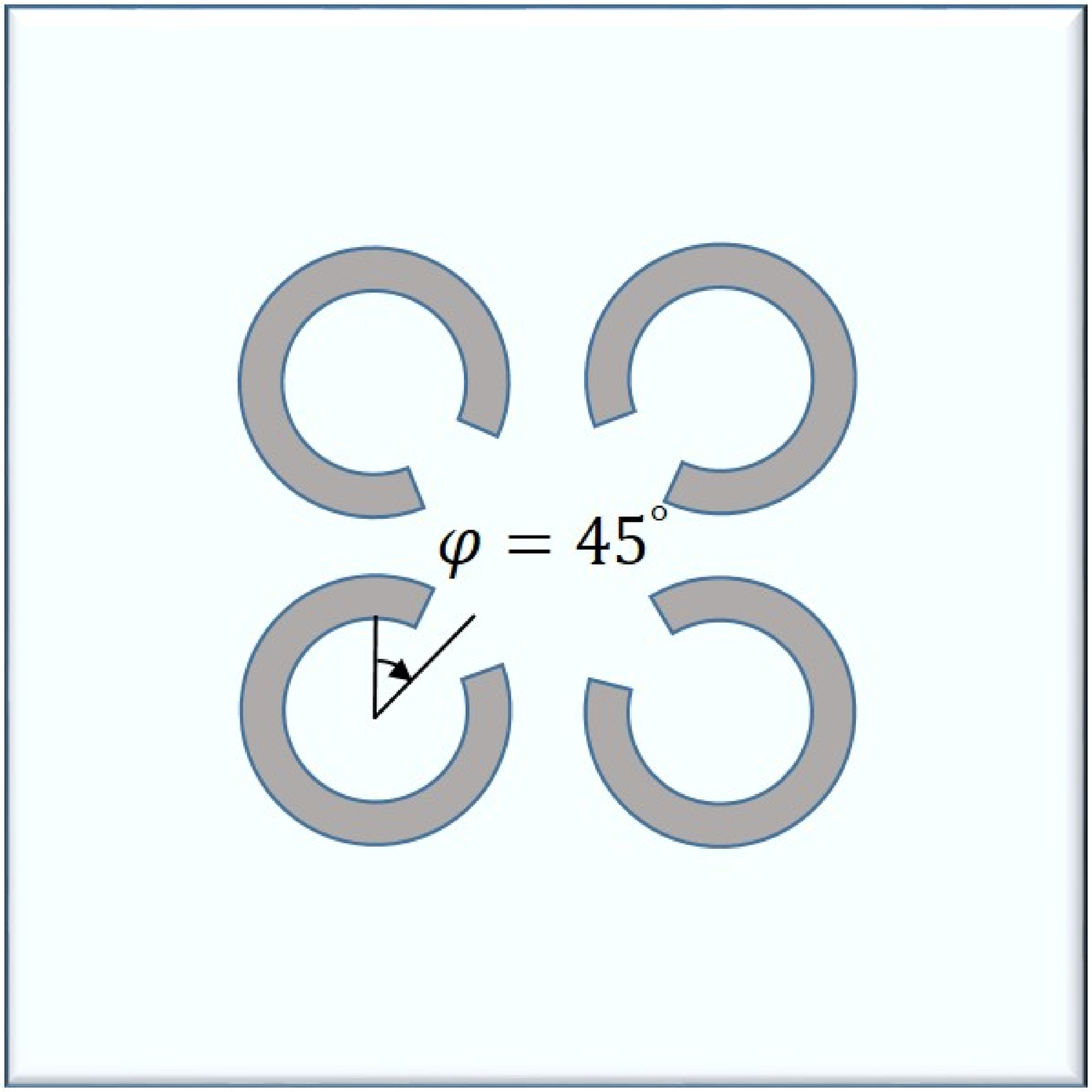}
}
\subfloat[]{
\includegraphics[scale=0.16]{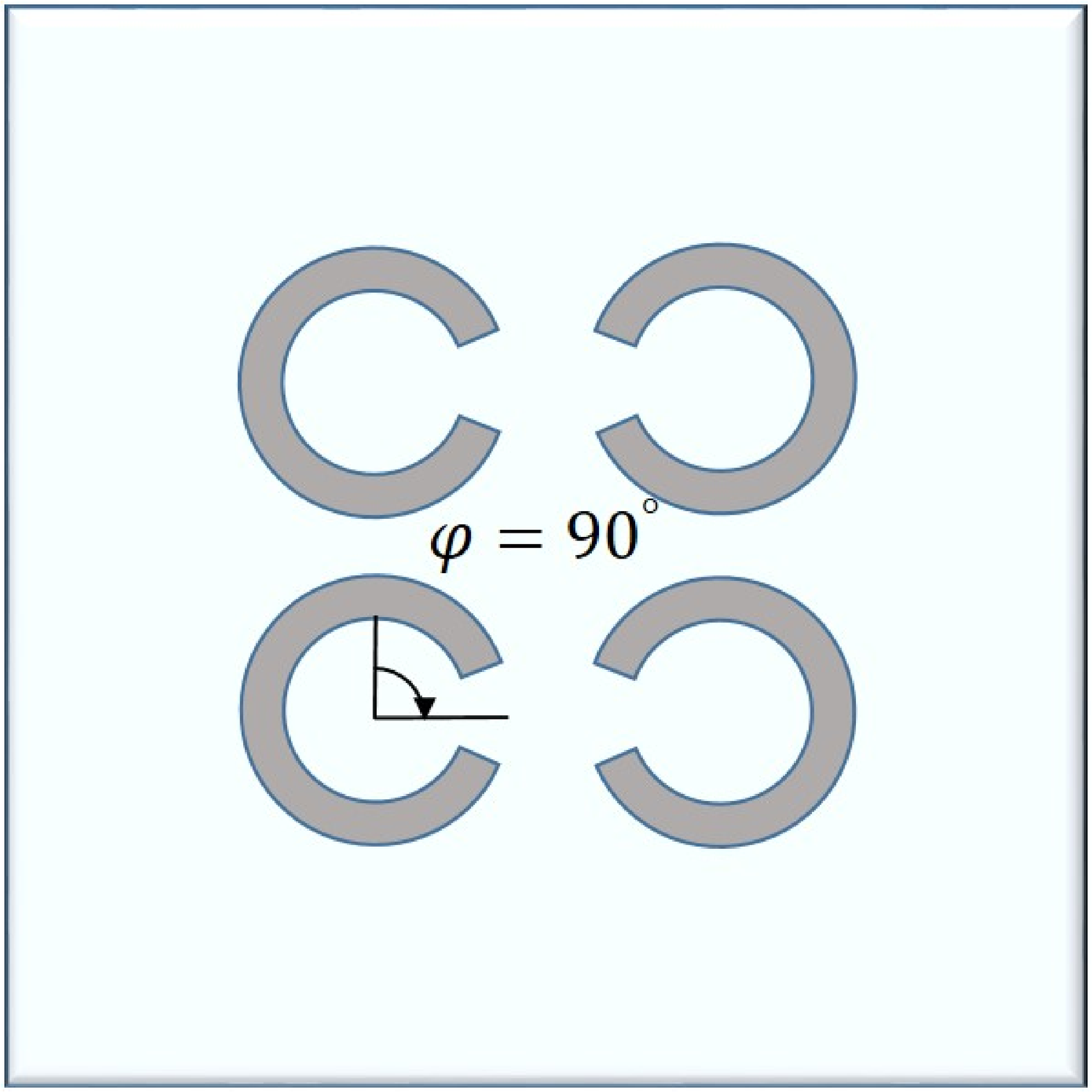}
}
\caption{Three quad-supercells of the SRRs with different mutual rotation angles (a) $0^{\circ}$, (b)
$45^{\circ}$, and (c) $90^{\circ}$.}
\end{figure}

\section{Simulation results and discussion}\label{sec3}
The simulated transmission spectra for the three different metamaterial arrays are shown in Fig. 2.
\begin{figure}[htbp]\label{fig:2}
\centering
\includegraphics[scale=0.4]{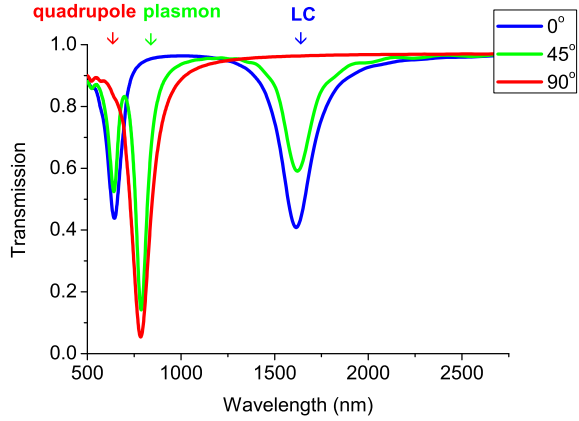}
\caption{Transmission spectra of metamaterial arrays with different mutual rotation angles: (a)$0^{\circ}$, (b)$45^{\circ}$ and (c)$90^{\circ}$.}
\end{figure}

As expected from the broken symmetry, the transmission spectra of the SRRs arrays show the strong polarization dependence. When the rotation angle is $0^{\circ}$ and the electric field $E$ is paralleled to the split, two resonance dips generated at 1615.19 nm and 644.628 nm are recognized as the $LC$ ($N=1$) mode and quadrupole ($N=3$) mode, respectively. However, when the rotation angle is $90^{\circ}$, only a single resonance dip is generated at 785.235 nm and it is identified as plasmon ($N=2$) mode. Amazing happens when the rotation angle turns to $45^{\circ}$, three resonance dips emerge at nearly the same wavelengths where $N=1$, $N=3$ and $N=2$ resonance dips previously located when the mutual rotation angle is $0^{\circ}$ and $90^{\circ}$, respectively, suggesting both the odd ($N=1$, $N=3$) and the even ($N=2$) modes are excited simultaneously, which is previously forbidden due to the symmetry constraints of orthogonally rotational SRRs arrays.

In order to get insights into the underlying physical mechanism, the electric field enhancement and surface charge density distributions at resonances in the z plane are plotted in Fig. 3 and 4.
\begin{figure}[htbp]\label{fig:3}
\centering
\subfloat[]{
\includegraphics[scale=0.2]{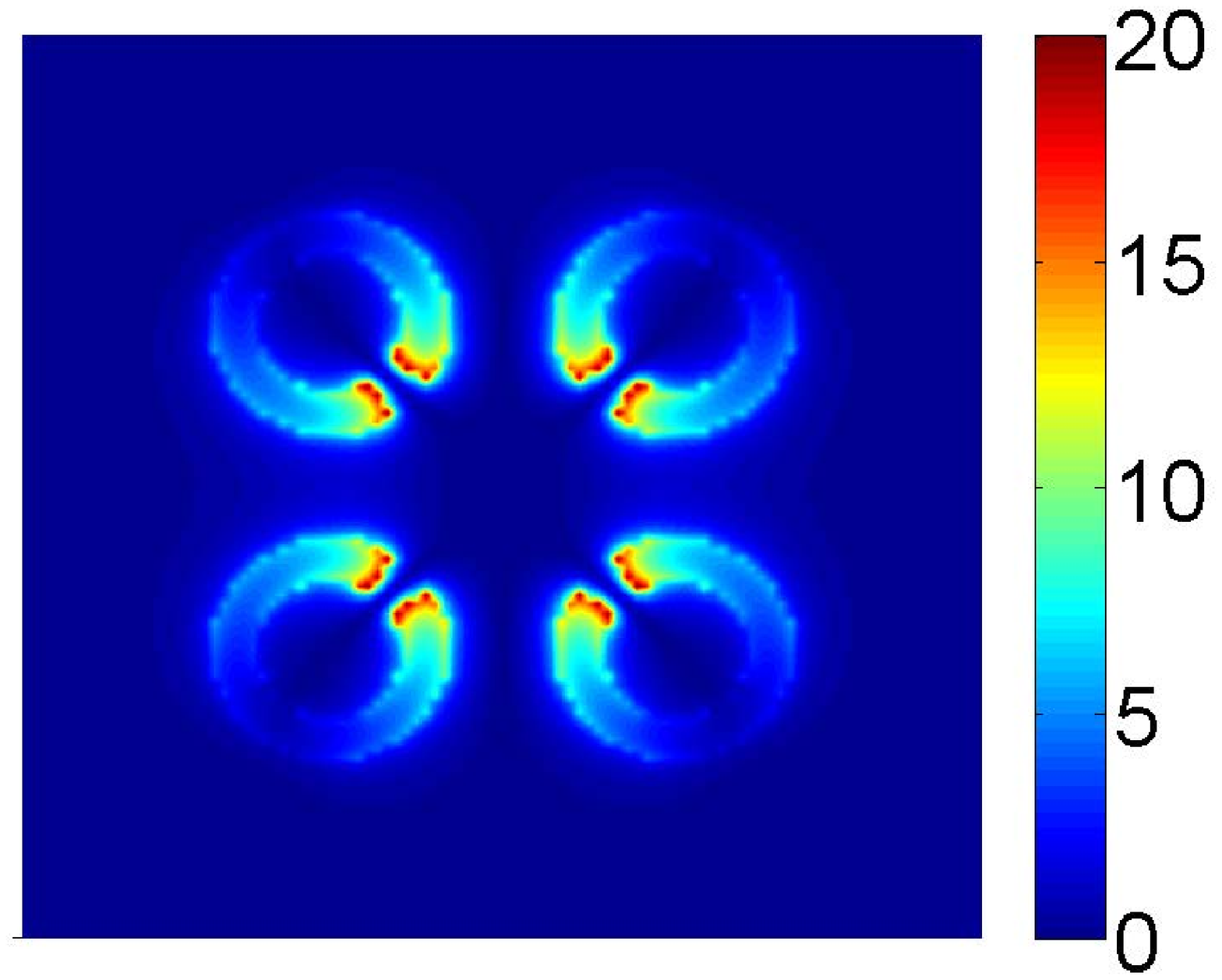}
}
\subfloat[]{
\includegraphics[scale=0.2]{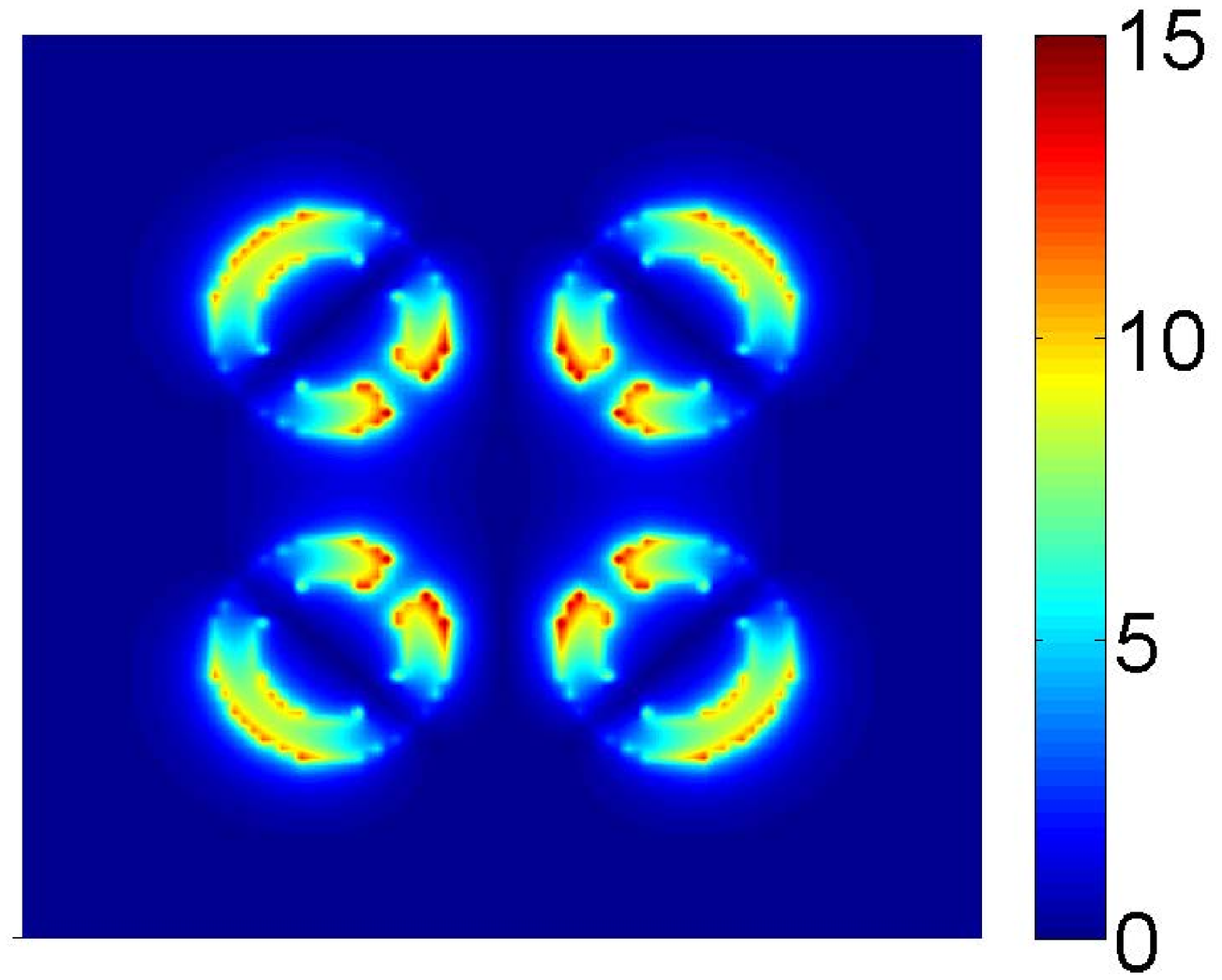}
}
\subfloat[]{
\includegraphics[scale=0.2]{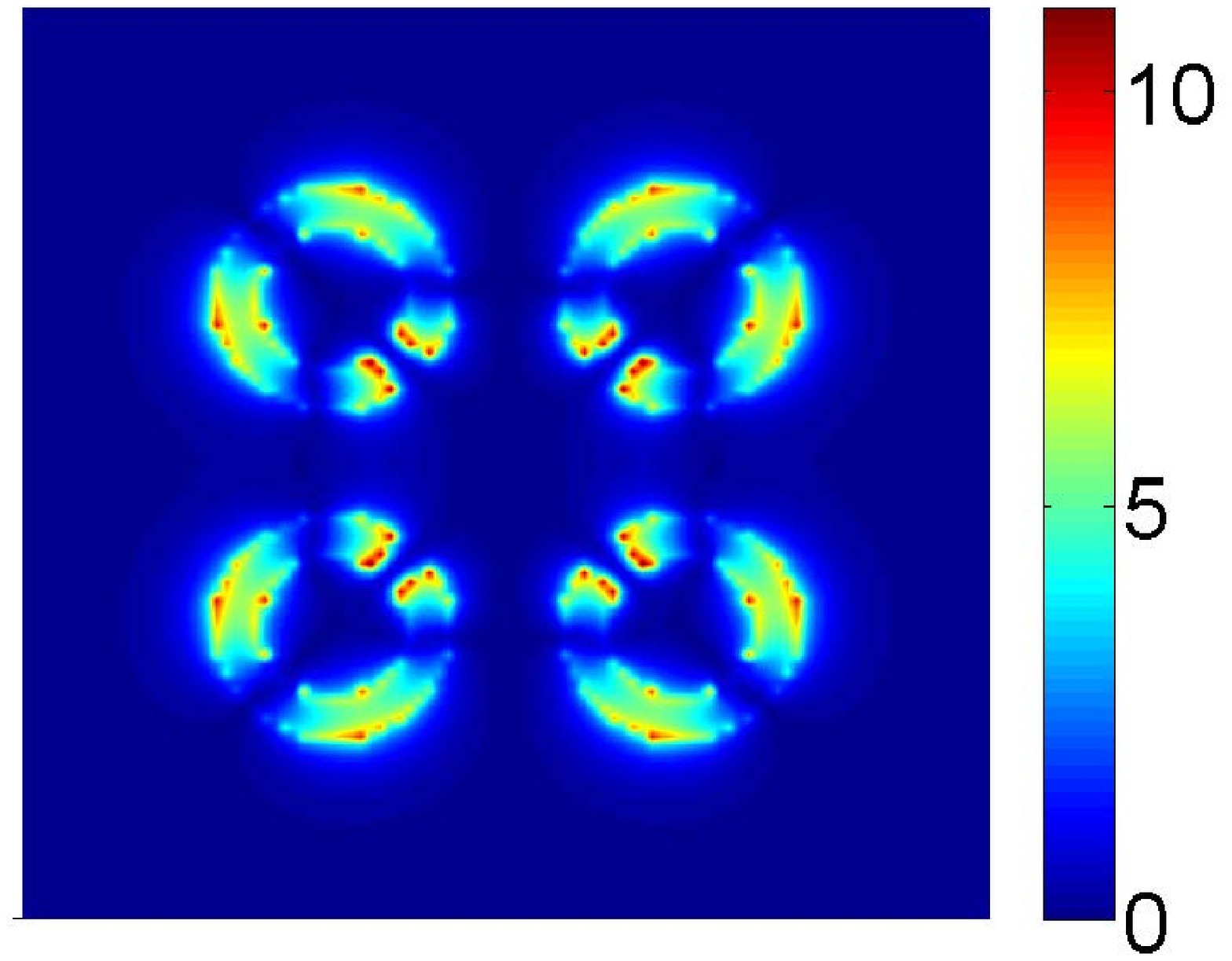}
}
\caption{Electric field enhancement at the angle of $45^{\circ}$ of (a)$LC$ mode ($N=1$), (b)plasmon mode ($N=2$) and (c)quadrupole mode ($N=3$).}
\end{figure}
\begin{figure}[htbp]\label{fig:4}
\centering
\subfloat[]{
\includegraphics[scale=0.2]{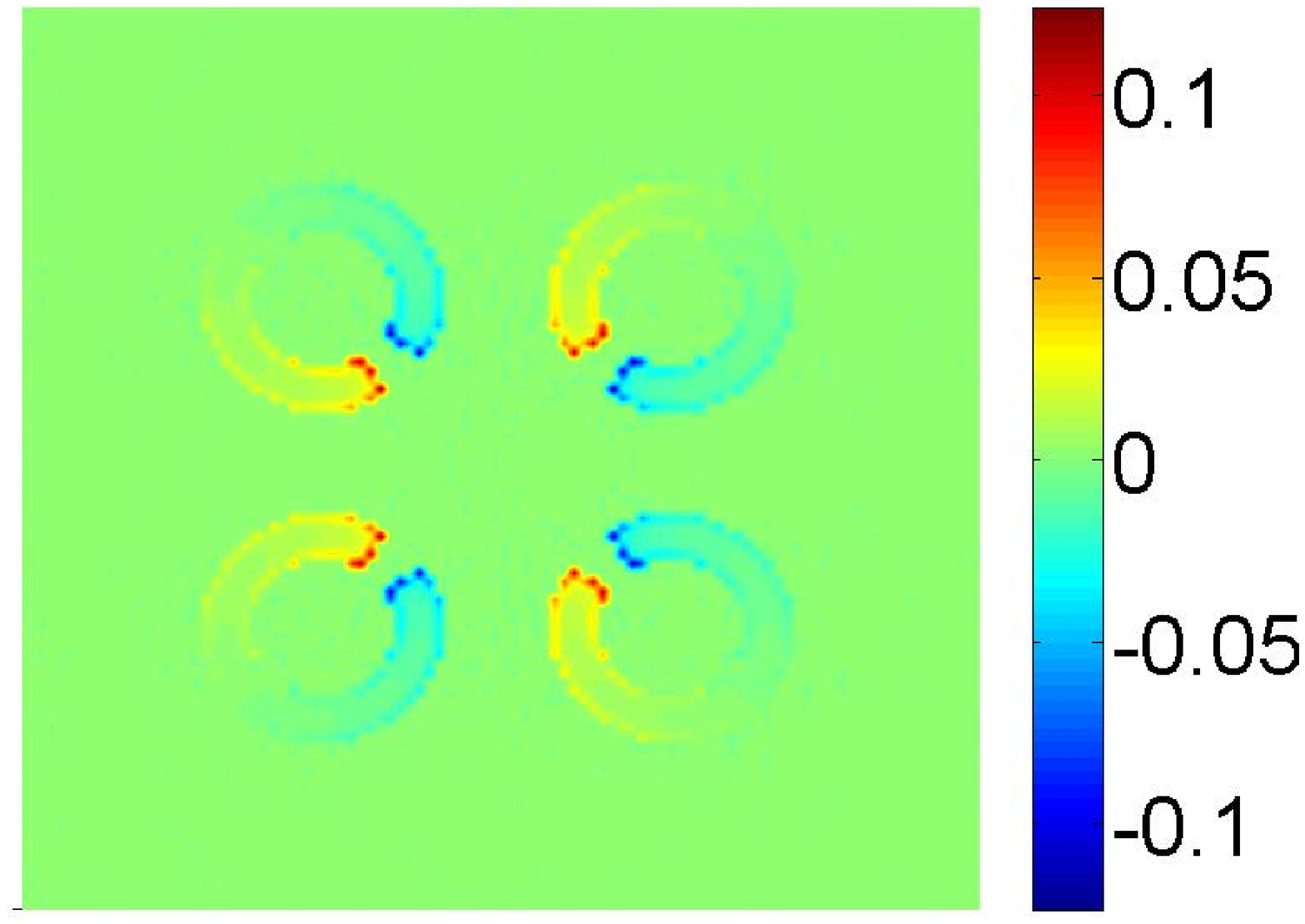}
}
\subfloat[]{
\includegraphics[scale=0.2]{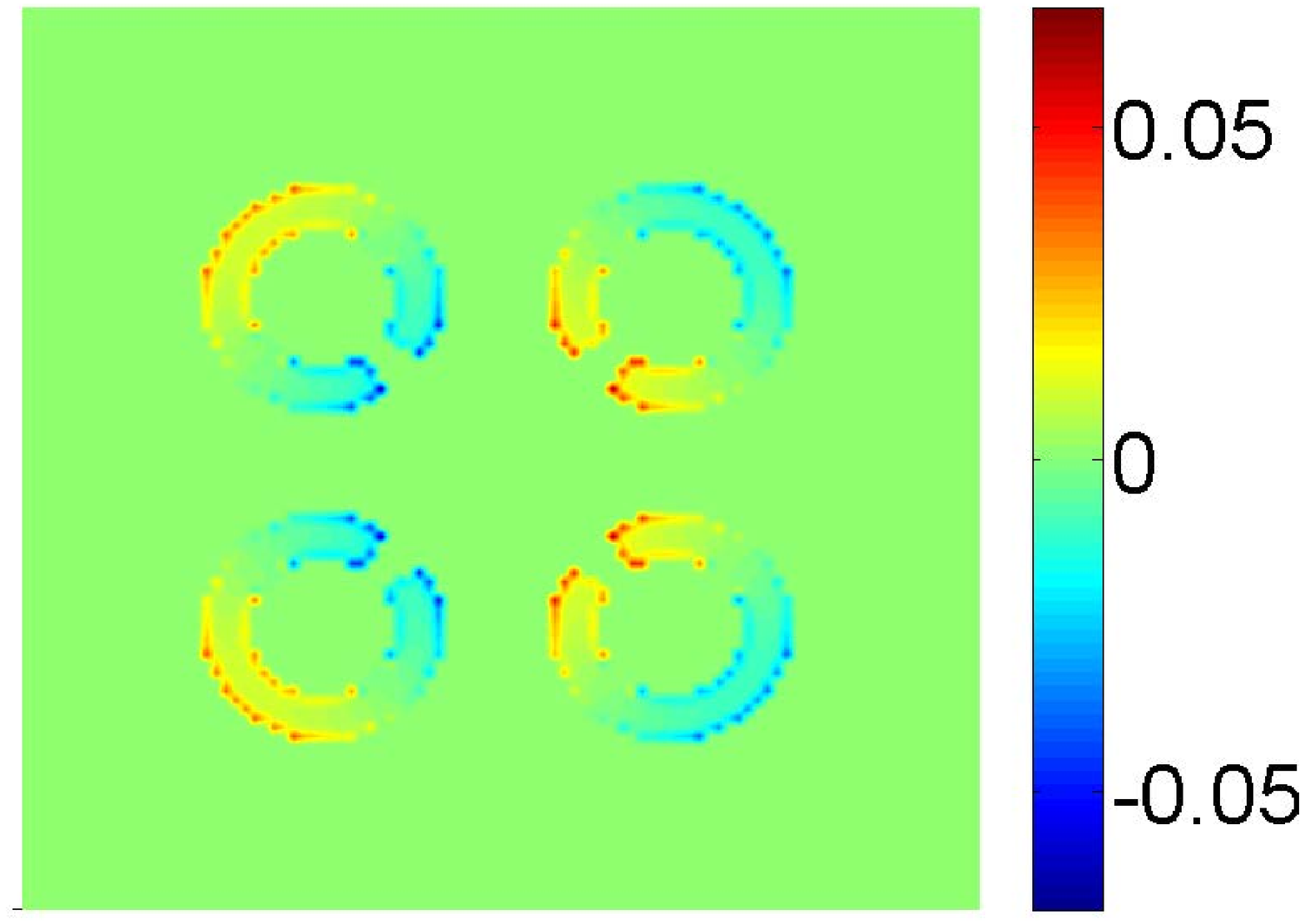}
}
\subfloat[]{
\includegraphics[scale=0.2]{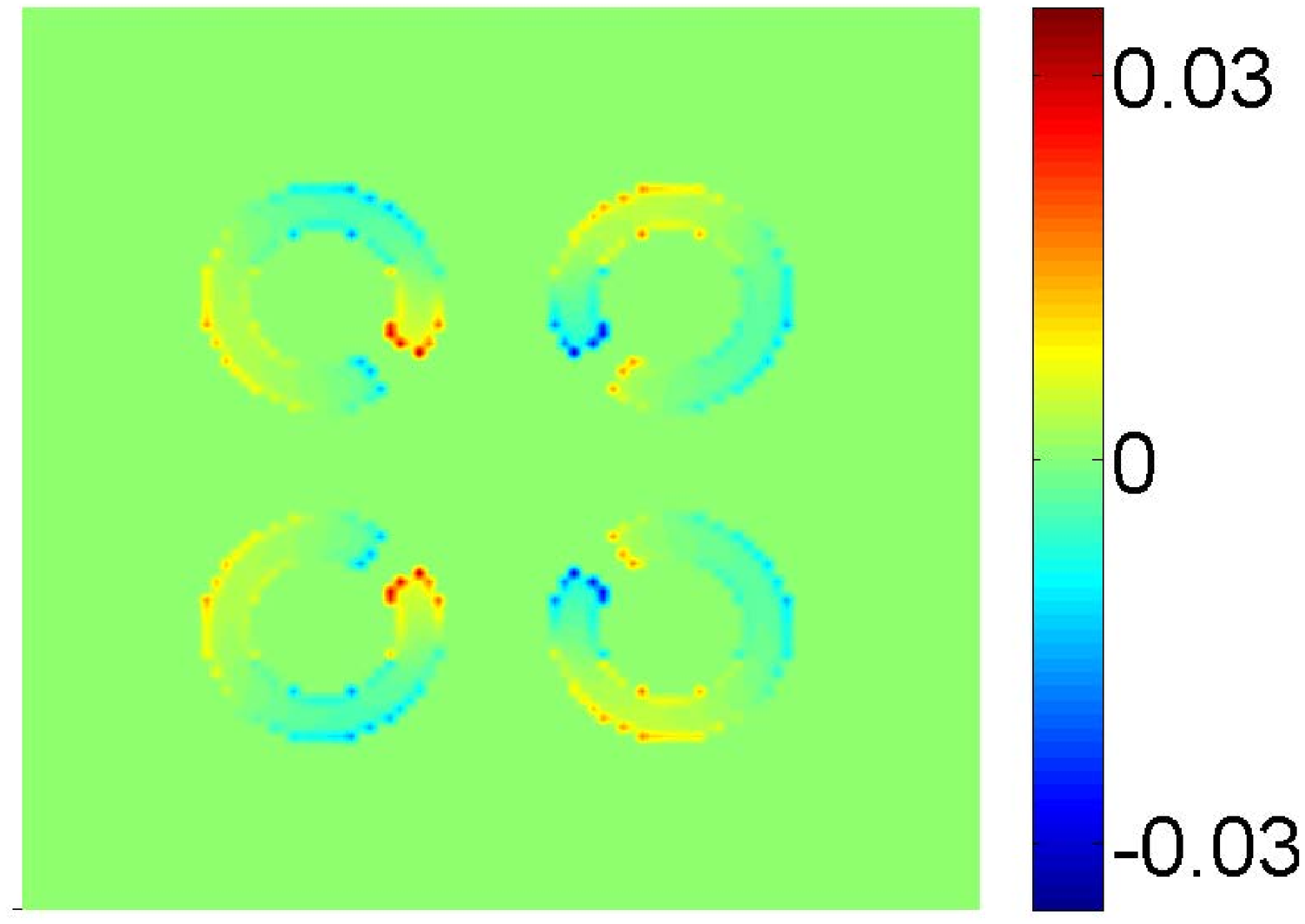}
}
\caption{Surface charge distribution at the angle of $45^{\circ}$ of (a)$LC$ mode ($N=1$), (b)plasmon mode ($N=2$) and (c)quadrupole mode ($N=3$).}
\end{figure}

As indicated in Ref\cite{clark2007}, the mode order is described as one minus the number of nodes, thus it is straightforward to confirm above-mentioned resonance modes show the characteristic behaviors of the $N=1$, $N=2$ and $N=3$ modes. As shown in Fig. 3(a) and Fig. 4(a), for the resonance dip at 1622.18 nm ($N=1$), the electric field is mainly confined in the vicinity of the split and a strong circulating current in each SRR is observed, which validates its support for magnetic dipole response and could be understood in terms of a $LC$-oscillator circuit with the split serving as a capacitor. A typical electric dipole response, i.e., the plasmon mode, at 786.885 nm ($N=2$) is clearly revealed in Fig. 3(b) and Fig. 4(b), the net induced dipoles in the quad-supercell exhibit a clear tendency to follow the polarization of the incoming planewave. And at the shortest resonance $\lambda=641.316$ nm ($N=3$), the quadrupole mode is also verified, and the electric field confinement as well as the surface charge density distributions shown in Fig. 3(c) and Fig. 4(c) are comparatively weak.

For the potential application in biochemical sensing, it is significant that these LSPRs should be tuned to the corresponding wavelength, i.e., the Vis-NIR portion of the spectra. To achieve this, the effects of the structure parameters of the supercell metamaterials need to be investigated.

Fig. 5(a) shows the variation in offset transmission spectra for the quad-supercell metamaterials with different split angles. When $\theta$ changes from $45^{\circ}$ to $15^{\circ}$, red shifts from 1523.19 nm to 1767.71 nm (for $LC$ mode), from 778.702 nm to 795.242 nm (for plasmon mode) and from 625.251 nm to 666.429 nm (for quadrupole mode) are observed. These phenomena are also directly reflected with the resonance wavelength as a function of split angle in Fig. 5(b) and that this parameter affect the $LC$ resonance most while leave the other two modes nearly untouched can be well understood that nonsymmetrical current flow caused by the split leads to an increase in capacitance in $LC$ circuit, and smaller split angle leads to a larger capacitance thus increases the resonance wavelength.
\begin{figure}[htbp]\label{fig:5}
\centering
\subfloat[]{
\includegraphics[scale=0.3]{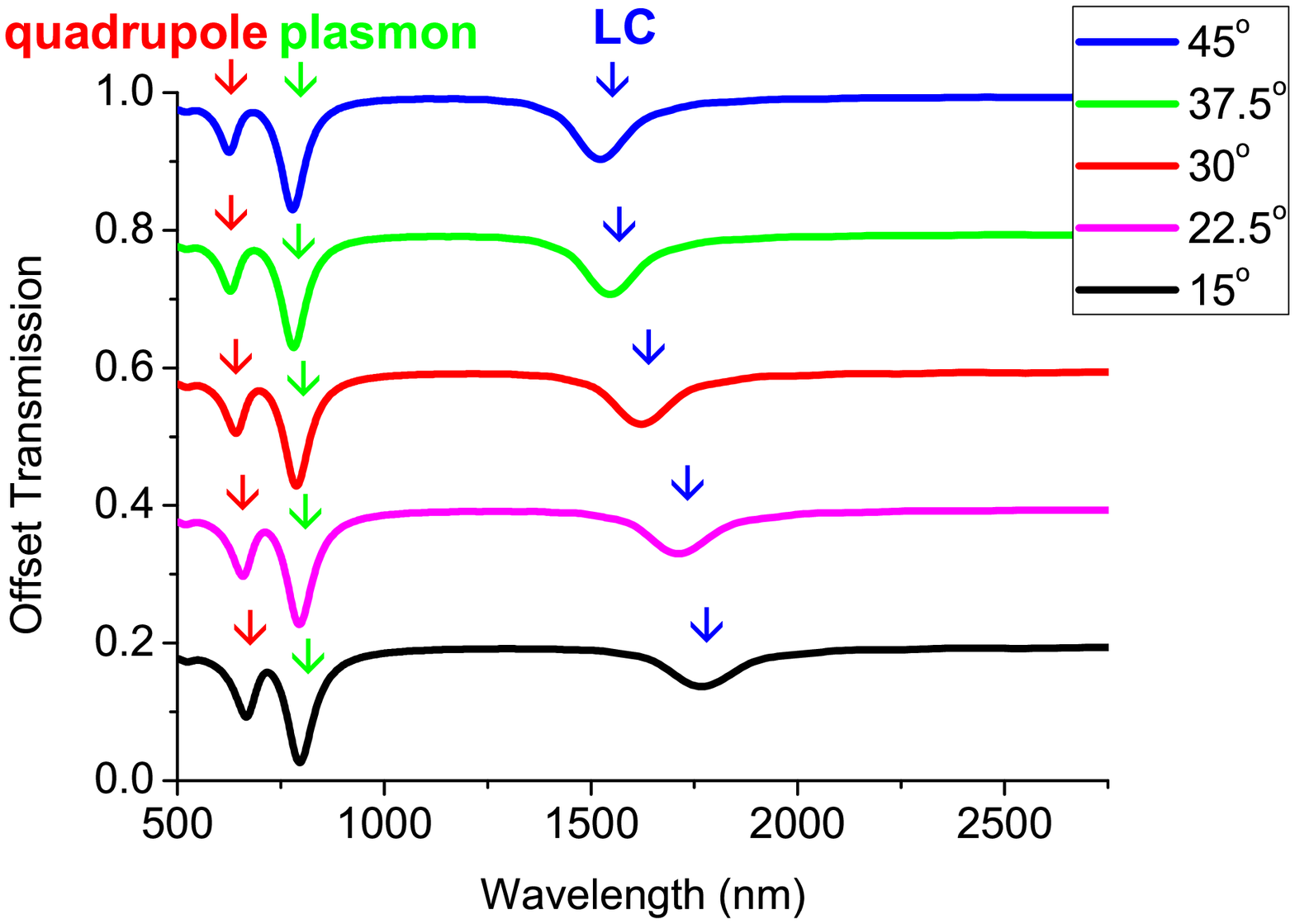}
}
\subfloat[]{
\includegraphics[scale=0.3]{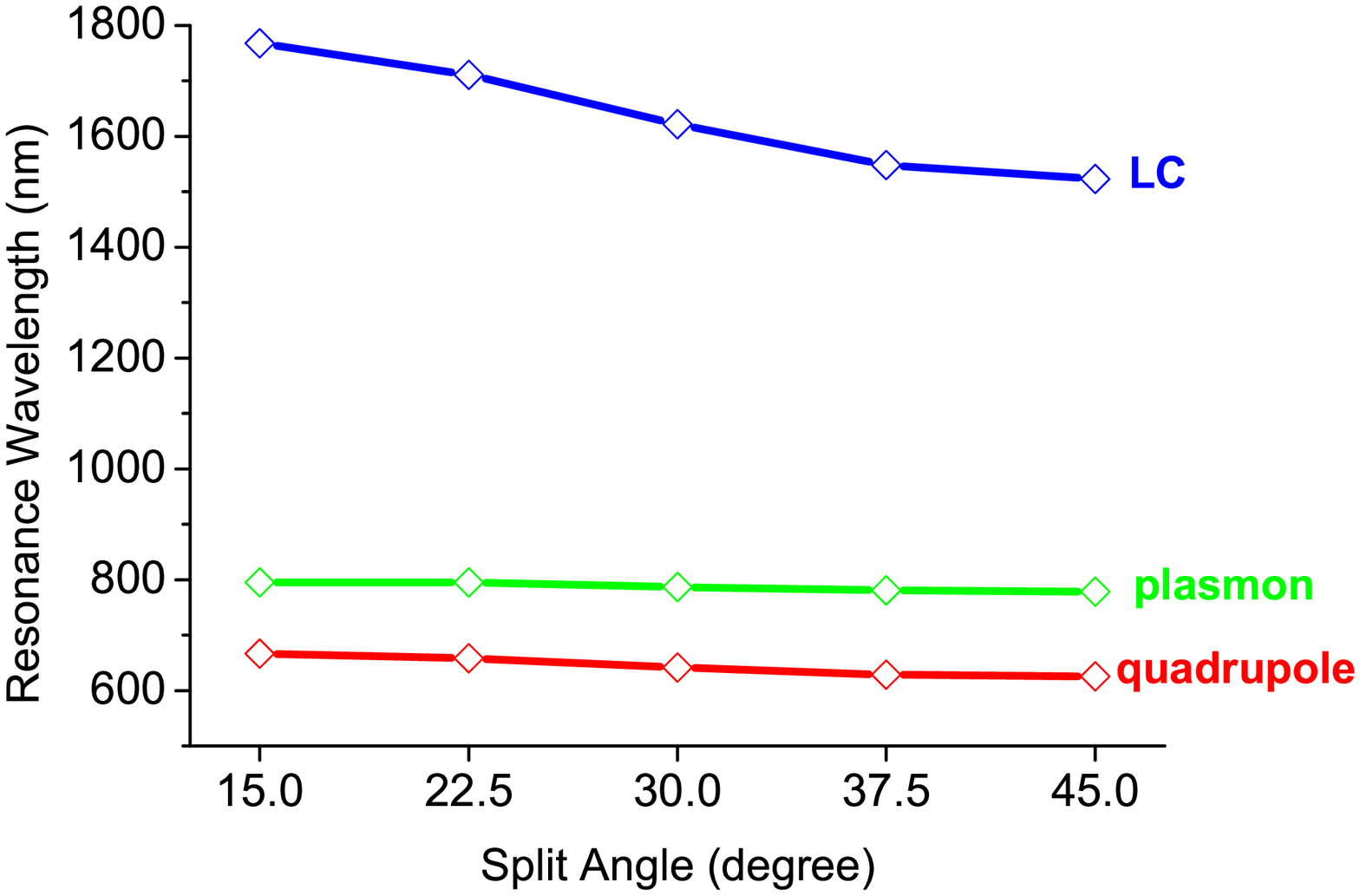}
}
\caption{(a)Offset transmission spectra and (b)resonance wavelengths for the quad-supercell metamaterials with different split angles.}
\end{figure}

In Fig. 6(a)-(c) the offset transmission spectra of the quad-supercell metamaterials as a function of the change in the size and lattice constants are presented. For $a=200$ nm, when the size of the SRR decreases from 140 nm to 60 nm, the $LC$ resonance exhibits clear blue shifts from 2678.11 nm to 1236.46 nm, the plasmon resonance from 1138.69 nm to 609.971 nm and the quadrupole resonance from 783.591 nm to 564.025 nm. For $a=180$ nm, as the SRR size decreases from $s=140$ nm to $s=60$ nm, blue shifts from 2129.69 nm to 1236.46 nm (for $LC$ mode), from 1015.18 nm to 609.971 nm (for plasmon mode) and from 783.591 nm to 563.177 nm (for quadrupole mode) are observed. And for $a=160$ nm, with the size ranging from 140 nm to 60 nm, the $LC$ resonance shifts from 2117.65 nm to 1236.46 nm, the plasmon resonance from 998.933 nm to 607.99 nm and the quadrupole resonance from 780.325 nm to 564.025 nm. An evident linear relation between the resonance wavelength and the size with each different lattice constant is clearly displayed in Fig. 6(d). The only notable exception arises when $s=140$ nm and $a=160$ nm, these four SRRs are tangent to each other and physically connected, thus the conductive coupling plays a significant role and leads to much strong coupling strength\cite{liu2010}. As a result, the $LC$ resonance shows a high redshift beyond NIR and we should abort this arrangement.
\begin{figure}[htbp]\label{fig:6}
\centering
\subfloat[]{
\includegraphics[scale=0.3]{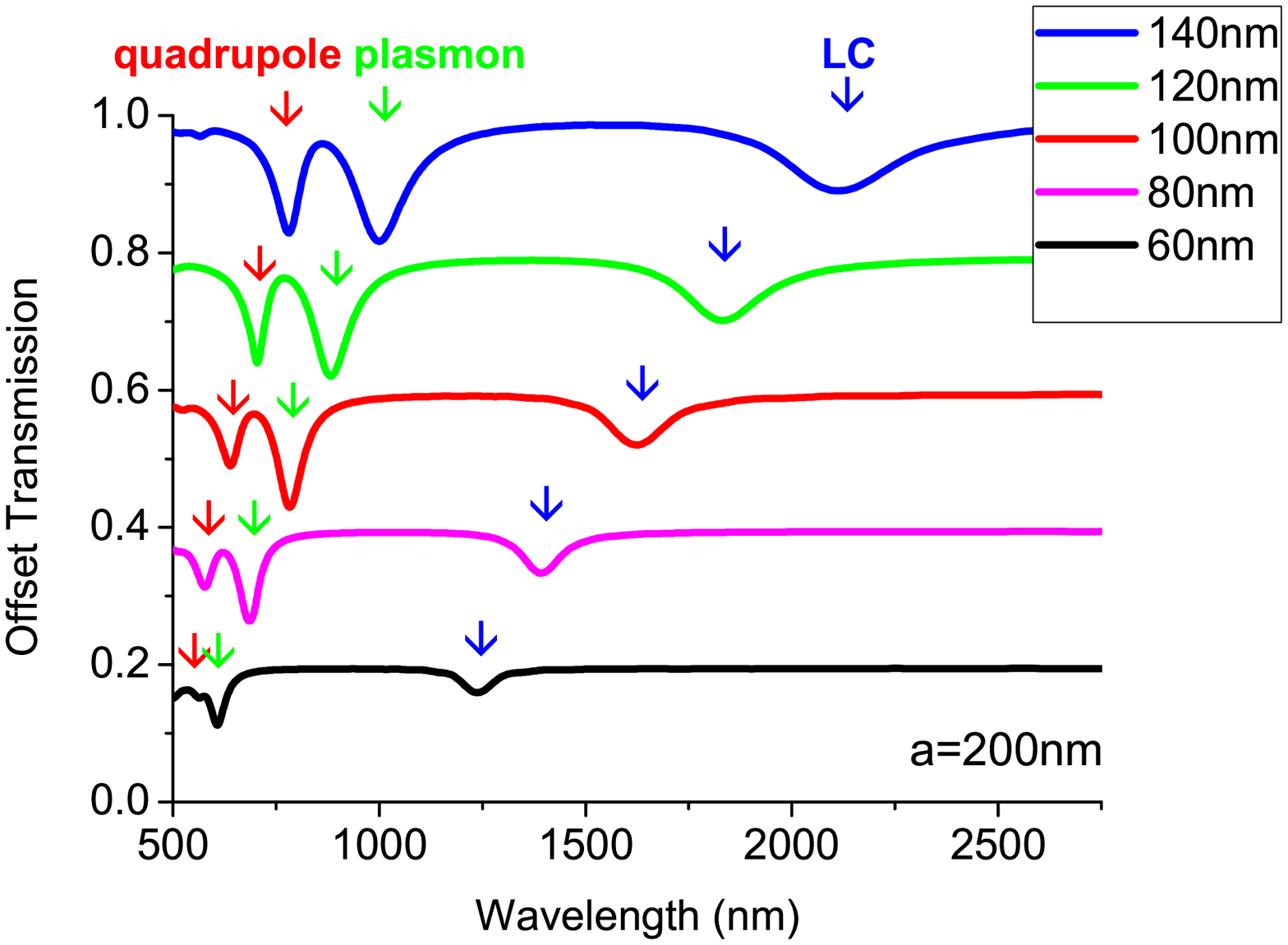}
}
\subfloat[]{
\includegraphics[scale=0.3]{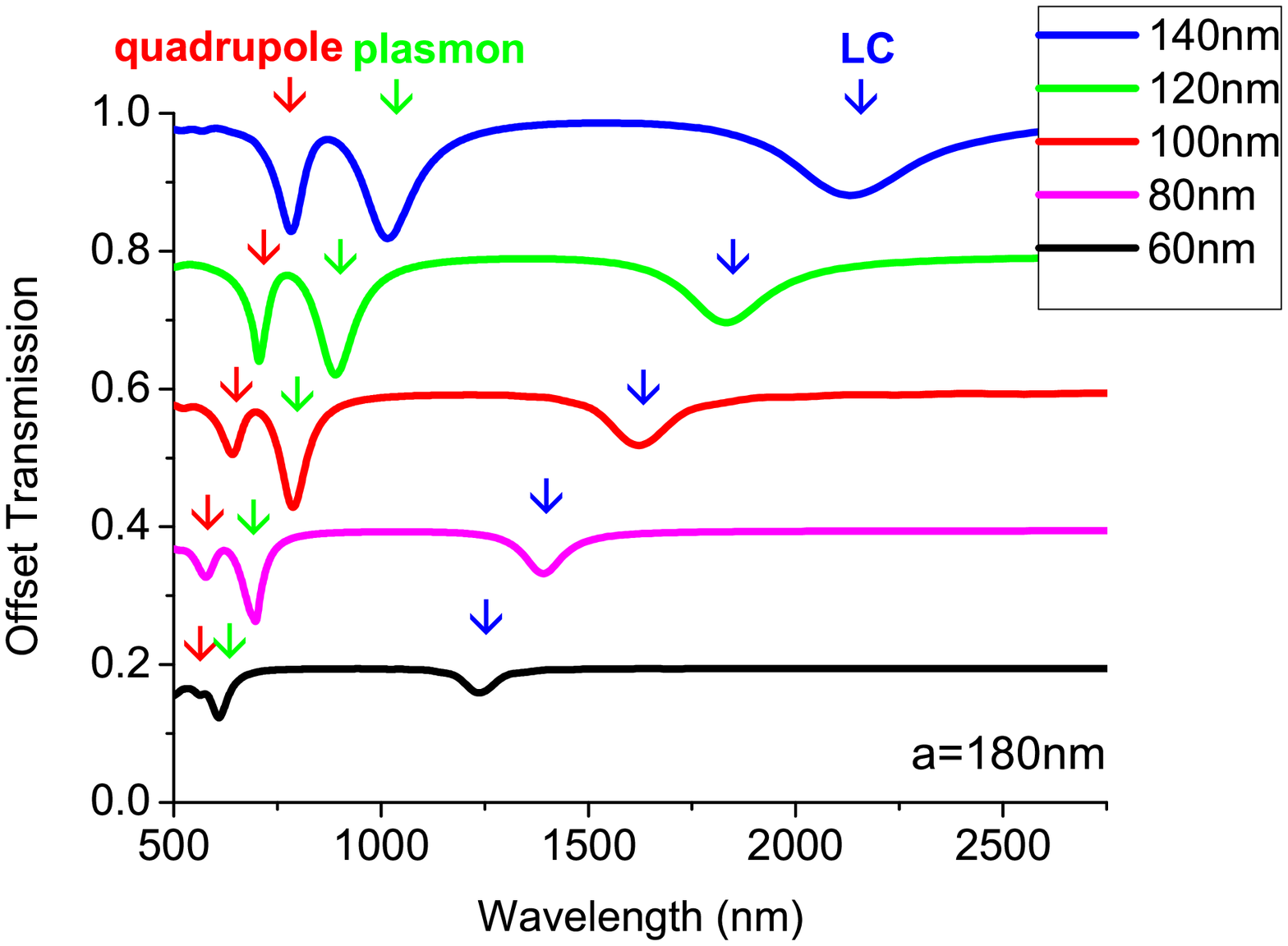}
}\\
\subfloat[]{
\includegraphics[scale=0.3]{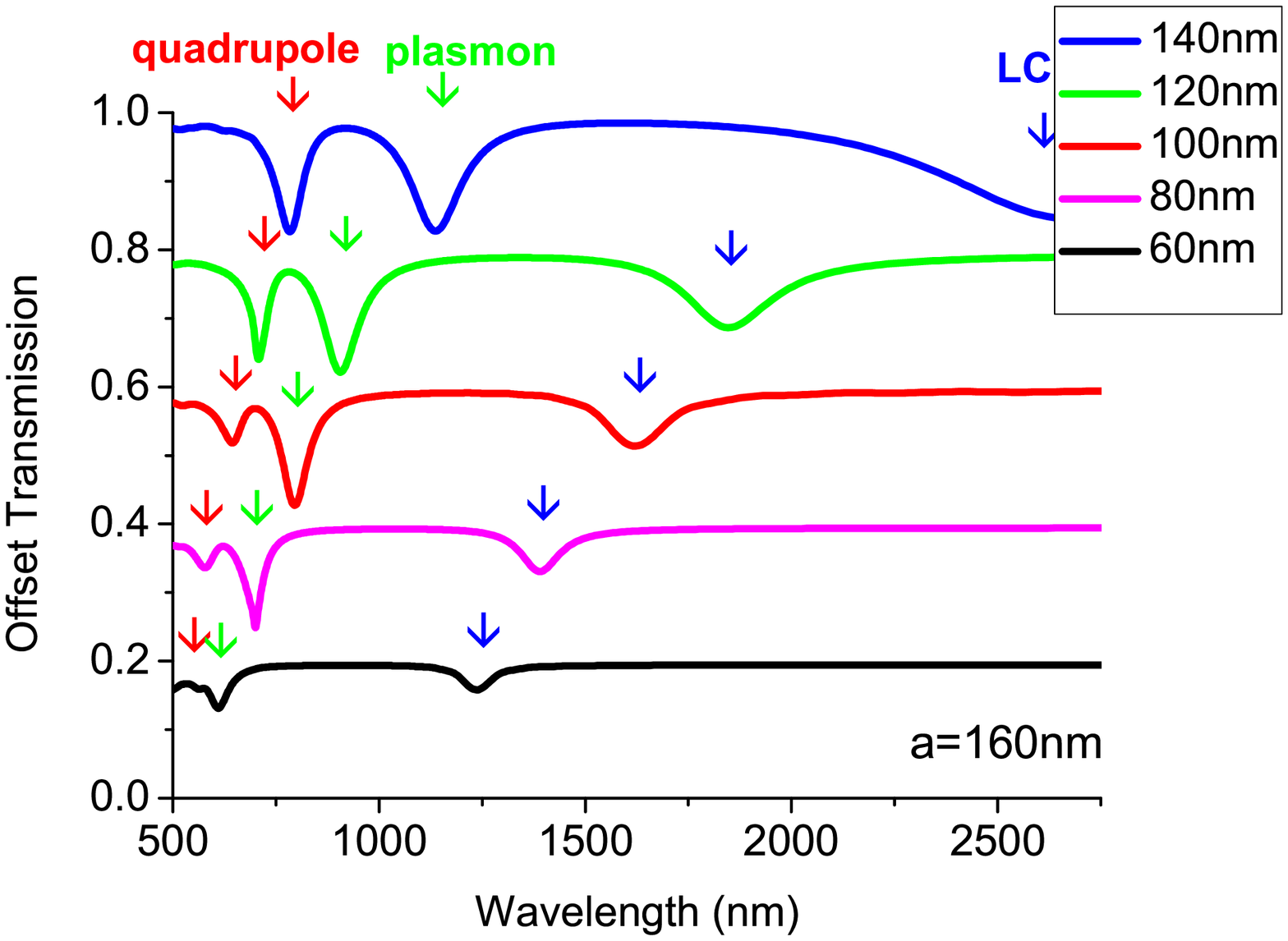}
}
\subfloat[]{
\includegraphics[scale=0.3]{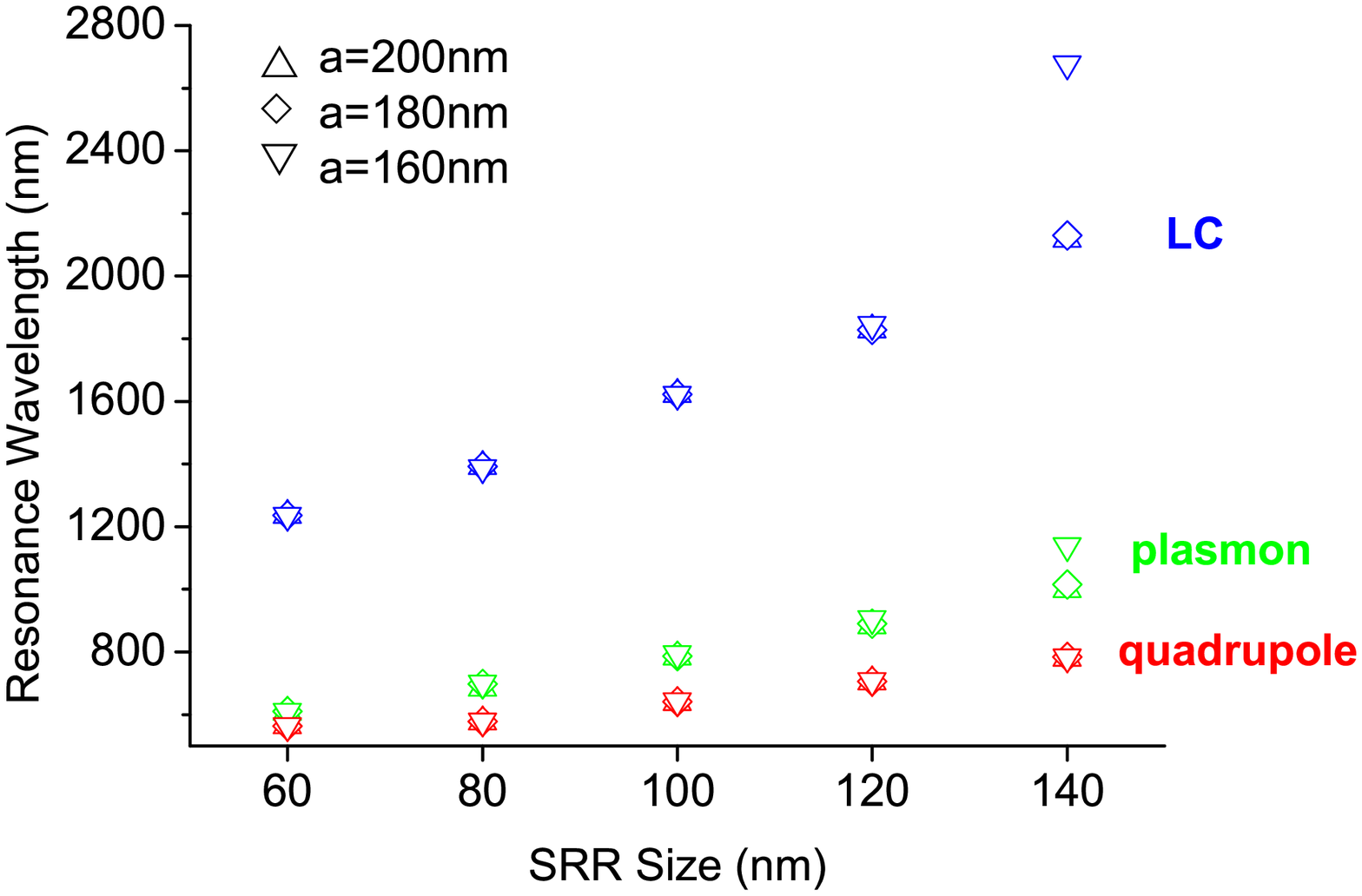}
}
\caption{(a)-(c)Offset transmission spectra and (d)resonance wavelengths for the quad-supercell metamaterials with different sizes and lattice constants.}
\end{figure}

Moreover, the simulations on the effects of the width and thickness have been also performed, and the results in Fig. 7 confirms that the SRRs exhibit longer resonance wavelength when the width (thickness) is narrower (smaller)\cite{rockstuhl2006apb,guo2007}.
\begin{figure}[htbp]\label{fig:7}
\centering
\subfloat[]{
\includegraphics[scale=0.3]{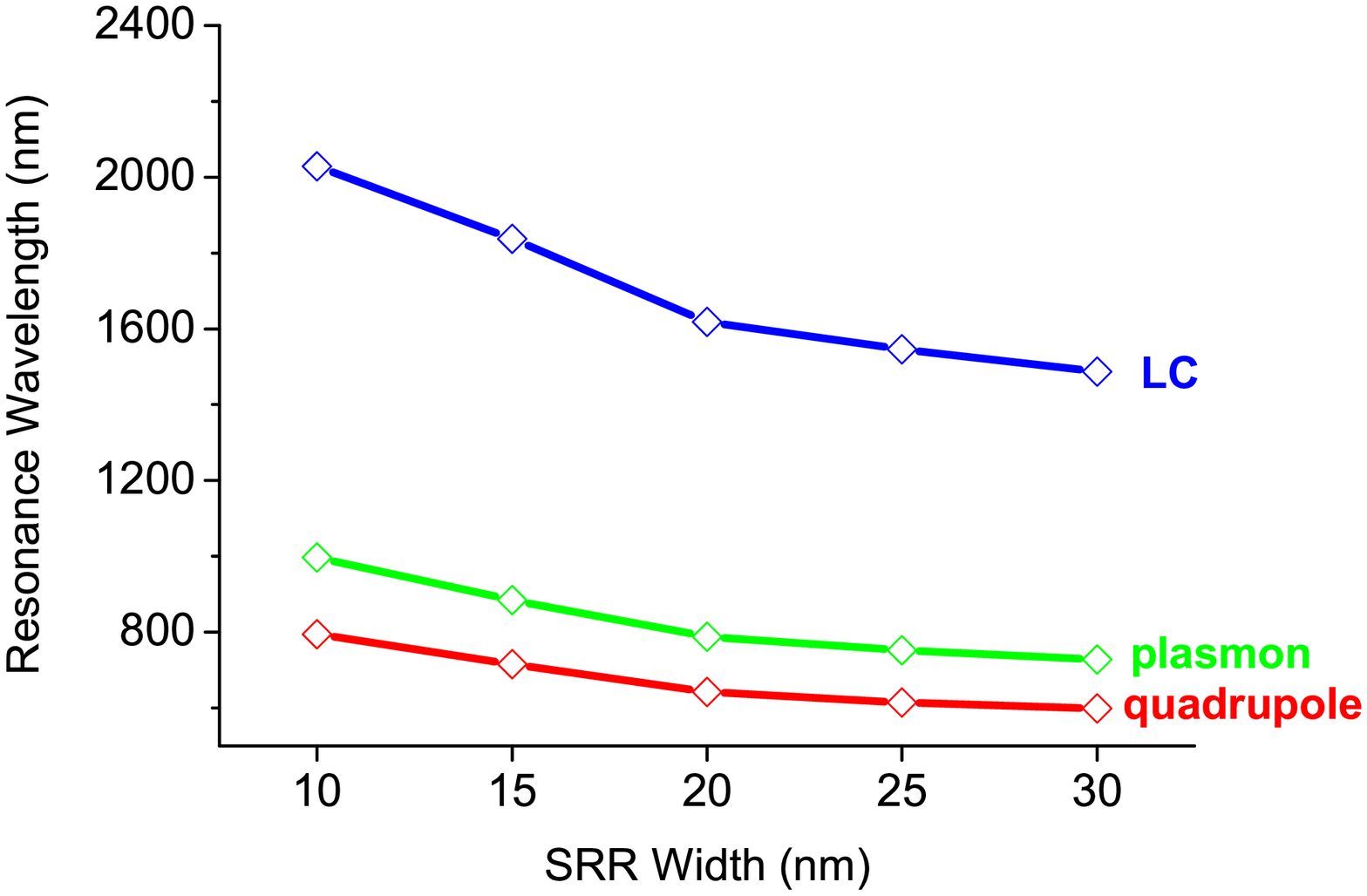}
}
\subfloat[]{
\includegraphics[scale=0.3]{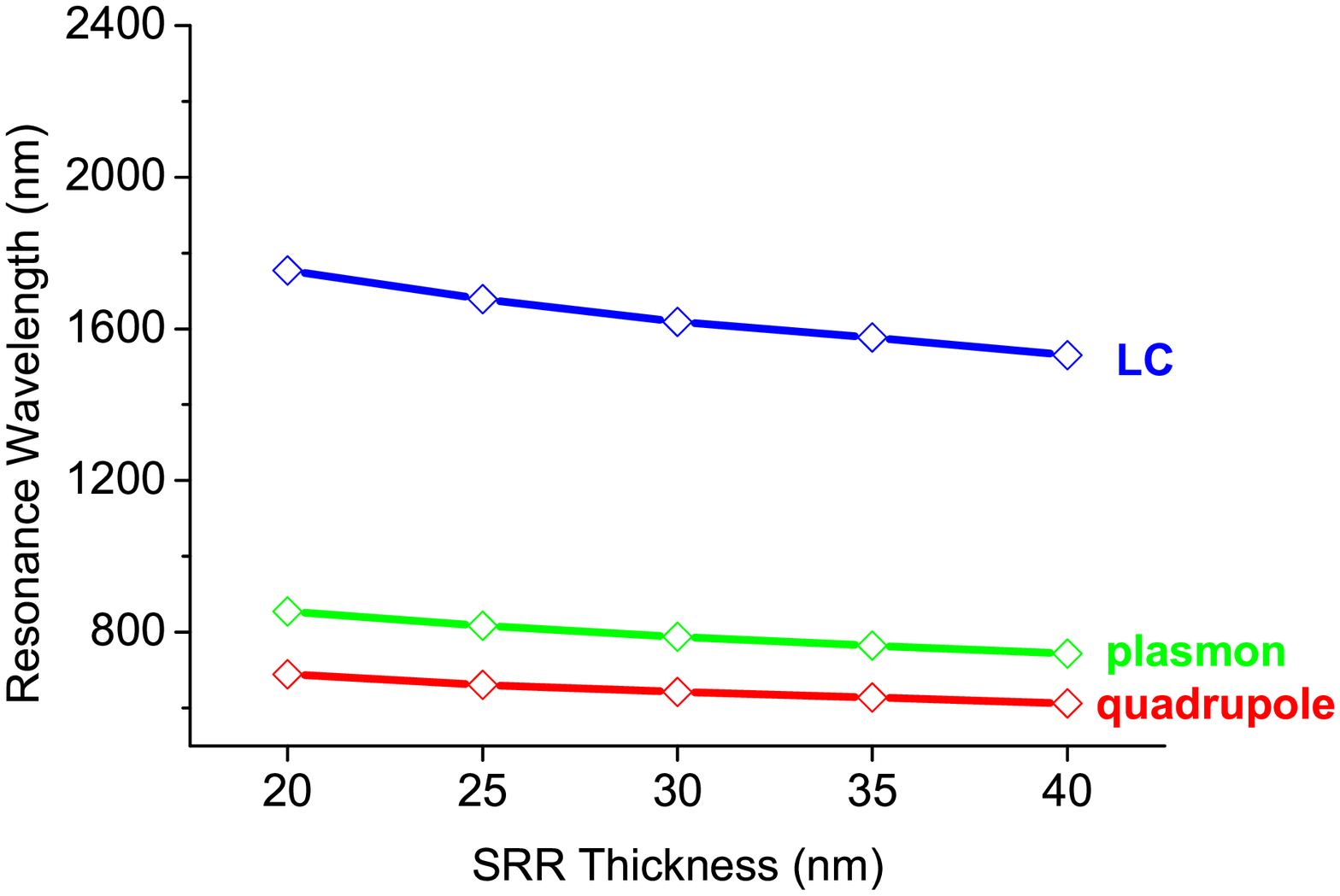}
}
\caption{Resonance wavelengths for the quad-supercell metamaterials with different (a)widths and (b)thicknesses.}
\end{figure}

It is found that the above shifts of the $LC$ mode and the plasmon mode almost cover the whole near infrared (NIR) spectrum, and these two resonance dips could be applied to multispectral sensing in NIR. The sensitivity is usually defined as the amount of peak(dip) shift per refractive index unit (RIU) change to characterize the performance of a SPR sensor\cite{tong2014}. In order to examine the sensitivity of our proposed RI sensor, the transmission spectra for the quad-supercell metamaterials with 100 nm thick cover layer filled with water ($n=1.332$) (blue) and $25\%$ aqueous glucose solution ($n=1.372$) (green) on the surface are respectively simulated. In Fig. 8, shifts of all the three resonance dips to longer wavelengths are clearly observed when the cover layer is filled with aqueous glucose solution instead of water due to the slight increase of the RI of surrounding environment. The wavelength shifts are 41 nm ($LC$ mode), 18 nm (plasmon mode) and 12 nm (quadrupole mode), which correspond to a set of sensitivities 1018 nm/RIU, 446 nm/RIU and 294 nm/RIU.
\begin{figure}[htbp]\label{fig:8}
\centering
\includegraphics[scale=0.4]{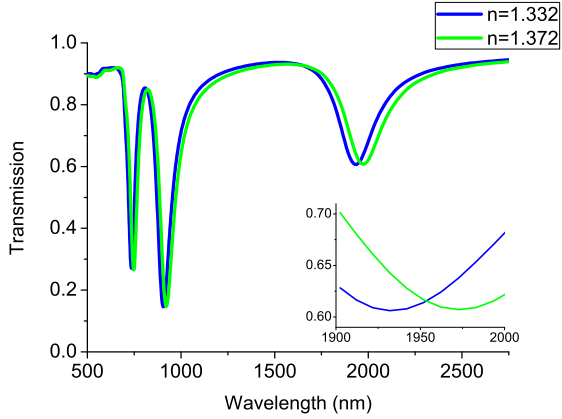}
\caption{The transmission spectra for the quad-supercell metamaterials by changing 100 nm thick cover layer filled with water ($n=1.332$) to $25\%$ aqueous glucose solution ($n=1.372$). Inset: enlarged figure of the shift of the transmission dip.}
\end{figure}

Now that the $LC$ resonance mode wavelength depends on the cover layer permittivity $(\epsilon_c=n_c^2)$ and the resonator size as $\lambda_{LC}\propto s(\epsilon_c)^{1/2}$, the sensitivity of $LC$ mode should have a positive correlation with the size of the SRRs $S=d\lambda_{LC}/dn_c\propto s$\cite{tobing2013}. Therefore the effects of the size on the sensitivity have been investigated, for the largest SRRs with $s=140$ nm, the simulated spectrum shifts are 52.49 nm ($LC$ mode) and 22.67 nm (plasmon mode), which correspond to 1312.25 nm/RIU and 566.75 nm/RIU, and for the smallest SRRs with $s=60$ nm, the shifts are 30.37 nm ($LC$ mode) and 9.23 nm (plasmon mode), which correspond to 759.25 nm/RIU and 232.13 nm/RIU. The sensitivity as a function of the size is plotted in Fig. 9, in the near infrared spectrum ($780\sim2526$ nm), the $LC$ mode holds a considerably high sensitivity from $\sim1500$ nm to $\sim2200$ nm and reaches the maximum 1017 nm/RIU within a 100 nm-SRR quad-supercell, on the other hand, the plasmon also possesses a good senstivity of 522 nm/RIU around 1000 nm within a 120 nm-SRR quad-supercell.
\begin{figure}[htbp]\label{fig:9}
\centering
\includegraphics[scale=0.4]{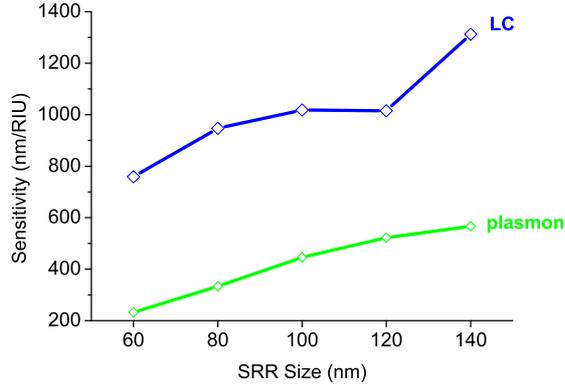}
\caption{Simulated sensitivity for the quad-supercell metamaterials with different sizes.}
\end{figure}

\section{Conclusion}\label{sec4}
In conclusion, the simultaneous excitations of both the odd ($N=1$ and $N=3$) and even ($N=2$) resonance modes in SRRs, which are forbidden due to symmetry constraints, have been realized in our proposed quad-supercell metamaterials, and the insensitivity to two orthogonal polarizations is automatically satisfied due to the mutual rotation angle of $45^{\circ}$. In order to make better use of these LSPRs, the effects of structure parameters have also been systematically investigated and phenomenologically explained. Ultrahigh sensitivities $\sim1000$ nm/RIU for $LC$ mode ($N=1$) and $\sim500$ nm/RIU for plasmon mode ($N=2$) are obtained in NIR spectrum and are expected to have bright prospects in ultrasensitive and multispectral biochemical sensing.
Furthermore, the design principles we have introduced in this work could be applied to other geometries of metamaterial sensors across large parts of the electromagnetic spectrum.

\begin{acknowledgements}
The author Shuyuan Xiao expresses his deepest gratitude to his Ph.D. advisor Tao Wang for providing guidance during this project and also thank Dr. Qi Lin (Hunan Univerisity) for beneficial discussion. This work is supported by the National Natural Science Foundation of China (Grant No. 61376055 and 61006045), and the Fundamental Research Funds for the Central Universities (HUST: 2016YXMS024).

\end{acknowledgements}



\end{document}